\documentclass{webofc}
\usepackage[varg]{txfonts}   
\usepackage{macros}
\usepackage{comment}
\usepackage{slashed}
\usepackage{tikz}
\usepackage{amsmath}
\begin{document}
\title{Gradient Flow: Perturbative and Non-Perturbative Renormalization}
%
%

\author{\firstname{Andrea} \lastname{Shindler}\inst{1}\fnsep\thanks{\email{shindler@frib.msu.edu}} 
}

\institute{
     Facility for Rare Isotope Beams, Physics Department, Michigan State University, East Lansing, Michigan, USA
}

\abstract{%
  We review the gradient flow for gauge and fermion fields and its 
  applications to lattice gauge theory computations. Using specific examples, 
  we discuss the interplay between perturbative and non-perturbative 
  calculations in the context of renormalization with the gradient flow.
}
\maketitle
\vspace{-1.0cm}
\section{Introduction}
\label{sec:intro}
In the last decade a new tool, 
called the gradient flow (GF)~\cite{Luscher:2010iy,Luscher:2011bx,Luscher:2013cpa}, 
has earned its place as one of 
the most interesting development in lattice QCD calculations.
The GF is a renormalizable ultraviolet smoothing procedure that  
modifies the short-distance behavior of fields. 
For lack of space it is not possible 
to review all the results obtained with the GF. 
The GF has a plethora of applications ranging from the 
definition of the topological charge density, the renormalization 
of gauge and fermion local fields, 
the running of the strong coupling and scale setting.
We mainly discuss results and applications that pertain to
the renormalization of local fields in lattice QCD
and the interplay between perturbative and non-perturbative
calculations. This is especially important for phenomenological
applications. While the gradient flow has been defined and studied also for 
other field theories
~\cite{Makino:2014sta,Makino:2014cxa,Suzuki:2015fka,Hieda:2017sqq,Kasai:2018koz}, 
in these proceedings we only focus on SU($3$) gauge theories.

The GF for gauge fields appeared for the first time 
in Ref.~\cite{Narayanan:2006rf} in the study of Wilson loops for 
SU($N$) gauge theories at large $N$. It was noticed that 
modifying the gauge links in the loop with an APE smeared 
link projected to SU($N$), the naive continuum limit 
was equivalent to have continuum gauge fields satisfying what we now 
call the GF equation for gauge fields.\footnote{Incidentally 
a local modification of the lattice static action using the same 
form of smearing was already suggested in Ref.~\cite{DellaMorte:2005nwx}
in the context of HQET.} In Ref.~\cite{Luscher:2009eq} the Wilson flow, a
discretized version of the GF, was introduced 
as a gauge field transformation 
in the context of trivializing maps for gauge theories, and 
in Ref.~\cite{Luscher:2010iy} the GF was presented as a new tool 
to address a set of lattice gauge theories calculations.
In particular it was shown~\cite{Luscher:2010iy,Luscher:2011bx} that despite
the apparent non-locality,  
correlation functions of flowed fields 
are still renormalizable to all orders in perturbation theory,
and the renormalizability extends to include fermion fields~\cite{Luscher:2013vga}.

In the first part of these proceedings we discuss the GF for gauge fields and results 
related to the topological charge, the strong coupling and the scale setting.
In the second part we present applications of the GF for fermions in the context
of renormalization of local fields. We conclude with a survey
of other results and final remarks.

\section{Gradient flow for gauge fields}
\label{sec:gf_gauge}
For gauge fields we consider the gradient flow defined by the following 
equation
\bea 
&&\partial_{t} B_\mu(x,t) = D_\nu G_{\nu\mu}(x,t)\,, \nonumber \\
&&\left.B_\mu(x,t)\right|_{t=0} = A_\mu(x)\,,
\label{eq:gf_gauge}
\eea 
where $A_\mu$ is the non-abelian gluon field, $B_\mu$ denotes the
flowed gluon field.
The flowed field tensor is defined as the unflowed one
$G_{\mu\nu} = \partial_\mu B_\nu - \partial_\nu B_\mu + \left[B_\mu,B_\nu \right] $,
and the flowed covariant derivative as $D_\nu = \partial_\nu + \left[B_\nu, \cdot \right]$.
It is immediate to notice that the flow time $t$ has energy dimension -2, thus the 
GF introduces a new length scale in the theory proportional to $\sqrt{t}$.
The GF equation at vanishing strong coupling
resembles a heat equation in $4$ dimensions and the solution is immediately found 
convoluting the heat kernel, $K(x,t)$ of the equation with the initial condition
\be 
B_\mu (x, t)= \int d^4 y~K(x-y, t) A_\mu(y)\,, \quad K(x, t)  = \int \frac{d^4p}{(2\pi)^4} {\textrm{e}}^{i p x } {\textrm{e}}^{-t p^2} = \frac{{\textrm{e}}^{-x^2/4t}}{(4\pi t)^2}\,.
\label{eq:tree}
\ee 
The effect of the GF is to perform a Gaussian damping on the large momenta modes of the
gauge fields, i.e. a smoothing a short distance over a range of $\sqrt{8t}$.
A key property of the flowed gauge fields is that
this tree-level result extends to all order in perturbation theory. 
L\"uscher and Weisz have shown~\cite{Luscher:2011bx} that the 
flowed gauge fields in correlation 
functions do not require any additional renormalization beside the
usual renormalization of the bare parameters of the theory. 

The presence of the flow time does not complicate much perturbative calculations,
and sophisticated technique already exists to extend 
the calculations to $2$-loops~\cite{Artz:2019bpr} and in same cases 
to $3$-loops~\cite{Harlander:2016vzb}. Numerically the GF can be integrated 
with infinitesimal "stout link smearing" steps~\cite{Morningstar:2003gk}, given the 
equivalence of the $2$ procedures.
\vspace{-0.3cm}
\subsection{Gradient flow and topological charge}
\label{ssec:topo}
\begin{figure}[h]
     \centering
     \includegraphics[width=6.3cm,clip]{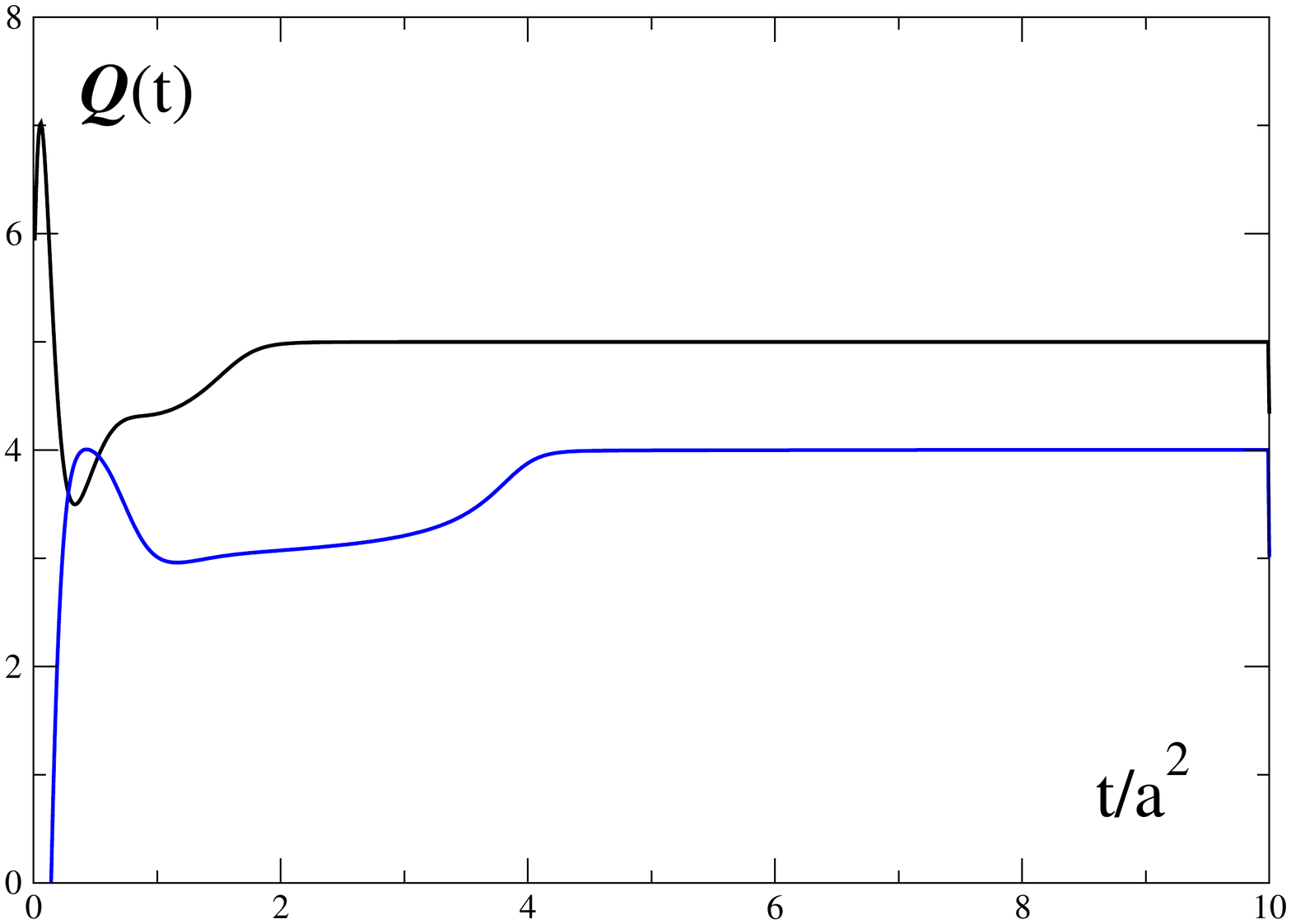}
     \includegraphics[width=6.3cm,clip]{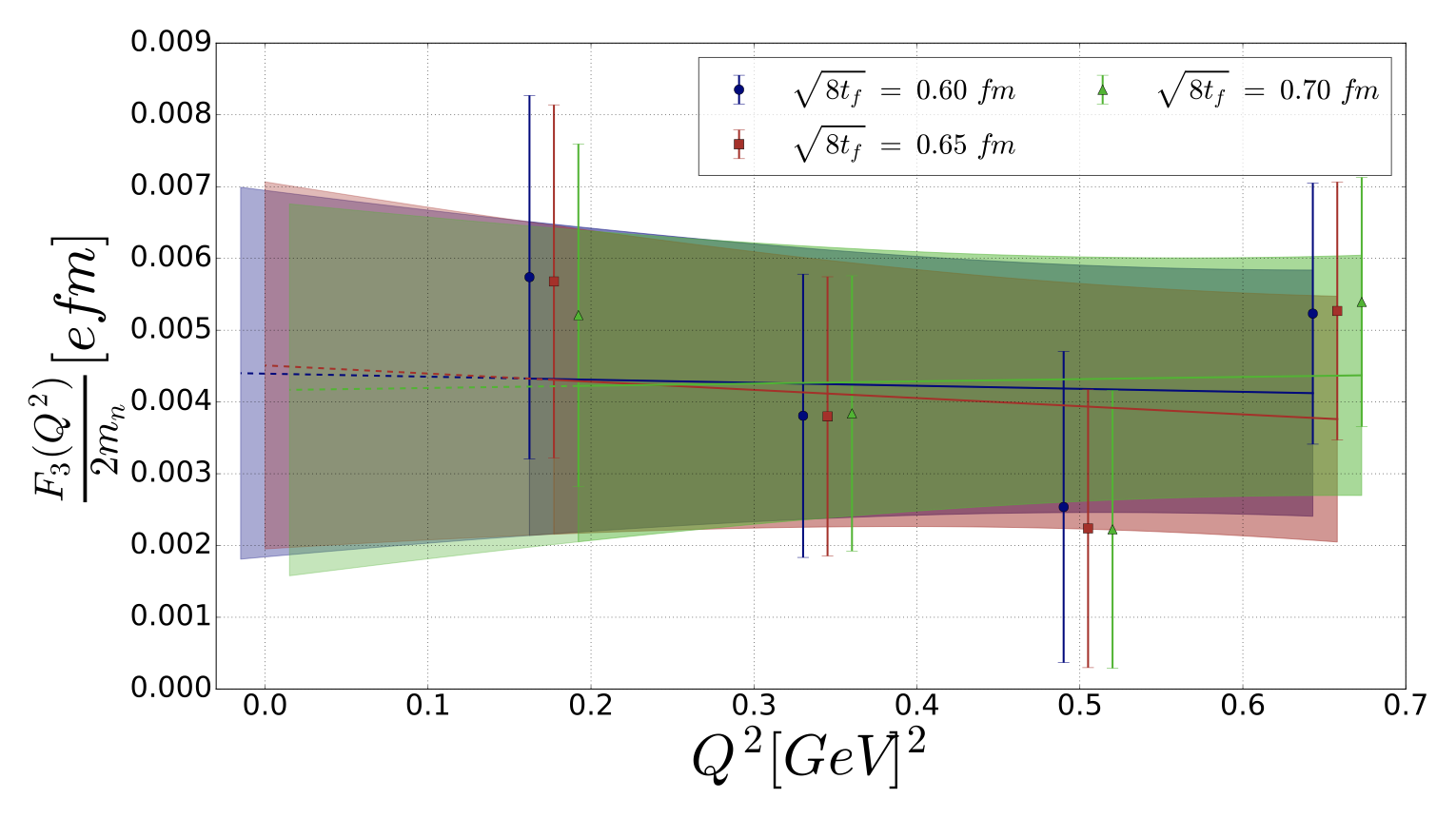}
     \caption{Left plot: Flow-time dependence of the topological charge on $2$ 
     SU($3$) Yang-Mills gauge configurations. 
     Right plot: Dependence of the nucleon CP-odd form factor 
     on the momentum transfer $Q^2$ for 
     a pion mass of $m_\pi \simeq 410$ MeV. 
     The different colors correspond to the different 
     flow times used to evaluate the topological charge.}
     \label{fig:topo_vs_flow} 
     \vspace{-0.5cm}      
\end{figure}

The GF for gauge fields provides a renormalizable,
and numerically easy to implement, definition of the topological 
charge $Q(t) = \int d^4x~ q(x,t)$ with topological charge density
\be 
q(x,t) = \frac{1}{32 \pi^2}\epsilon_{\mu\nu\rho\sigma}
\textrm{Tr}\left\{G_{\mu\nu}(x,t) G_{\rho\sigma}(x,t)\right\}\,.
\label{eq:gf_topo}
\ee 
The r.h.s of the GF equation~\eqref{eq:gf_gauge} can be written as the negative
derivative of the Yang-Mills classical action (evaluated on flowed gauge fields)
with respect to the field itself,
i.e. $-\frac{\delta S_{\mathrm{YM}}(B)}{\delta B_\mu(x,t)}$. This implies
that the GF equation naturally drives the fields towards the local minima 
of the theory~\cite{Luscher:2010iy}. 
This behavior can be observed in the left plot of 
Fig.~\ref{fig:topo_vs_flow},
from Ref.~\cite{Shindler:2015aqa}, where it is shown the flow-time 
dependence of the topological charge evaluated 
on 2 representative SU($3$) Yang-Mills gauge configurations.
We observe that after flowing to distances of $\sqrt{8t} \simeq 0.4 - 0.6$ fm 
the charge becomes independent of the flow time reaching a value 
very close to an integer. 
The property that the topological charge is independent of the flow time
is a result of the topological nature of the 
observable~\cite{Polyakov:1987ez,Ce:2015qha,Luscher:2021ygc}.
It can also be shown
that this definition of the topological charge is equivalent to the fermionic
definition based on Ginsparg-Wilson lattice Dirac operators in the Yang-Mills 
theory~\cite{Ce:2015qha} and in QCD~\cite{Luscher:2021ygc}. In conclusion 
the GF provides a computationally cheap and theoretically very robust 
definition of the topological charge. 

There is one caveat related to the potential large O($a^2$) cutoff effects 
that this definition might have when evaluated on the lattice. 
Tree-level improvements~\cite{Fodor:2014cpa}, improved lattice definitions
of the charge~\cite{BilsonThompson:2002jk} or improving 
flowed observables~\cite{Ramos:2015baa} following a more 
systematic Symanzik improvement program 
can help in performing the continuum limit.

In practice to define the topological charge in any correlation 
function it is sufficient to calculate it at flow times large enough 
as to avoid cutoff effects and not too large to avoid that the smoothing 
becomes too large and the correlators suffer from finite size effect.
Fixing the flow time in physical units one can then perform the continuum limit.
Examples are the calculation of the topological 
susceptibility of Refs.~\cite{Bruno:2014ova,Taniguchi:2016tjc}.
\vspace{-0.3cm} 
\subsection{Electric dipole moment from the $\theta$ term}
\label{sec:edm_theta}
\vspace{-0.1cm}
The very attractive properties of the GF definition of the topological charge
can be used to define other observables. One example is the electric dipole moment 
(EDM) induced by the $\theta$ term of QCD. 
In Euclidean space the theta term in the action
is given by $i \thetabar \int d^4x~\frac{1}{32 \pi^2}\epsilon_{\mu\nu\rho\sigma}
\textrm{Tr}\left\{G_{\mu\nu}(x) G_{\rho\sigma}(x)\right\}$.\footnote{
     The coefficient of the topological charge density is denoted by $\thetabar$
     to specify the particular choice where the mass matrix of the theory is real.} 
     The EDM of a nucleon, $d_{\mathrm{N}}$, is proportional
to the CP-odd form factor $F_3(Q^2)$, 
$|d_{\mathrm{N}}| = F_3(Q^2=0)/2 M_{\mathrm{N}}$, where $M_{\mathrm{N}}$ 
is the nucleon mass,
parametrizing the $Q^2$ dependence of the 
matrix element 
\be 
\left\langle N({\bf{p'}},s')|J_\mu^{\textrm{em}}|N({\bf{p}},s) \right\rangle =
\overline{u}_N({\bf{p'}},s') \Gamma_\mu^{\bar\theta}(Q^2)
u_N({\bf{p}},s)\,,
\label{eq:me}
\ee
evaluated in background where $\thetabar \neq 0$, and
where $\Gamma_\mu^{\bar\theta}(Q^2)$ is decomposed in CP-even and CP-odd form factors.
The Euclidean theory with $\thetabar \neq 0$ has a complex action, thus 
one way to calculate the matrix element~\eqref{eq:me} 
is to evaluate the correlation function 
\be 
\left\langle N(y_0,{\bf{p_2}}) J_\mu^{\mathrm{em}}(x_0,{\bf{q}}) 
\overline{N}(0,{\bf{p_1}})\right\rangle_{\thetabar}
\label{eq:corr_edm}
\ee
and expand in powers of $\thetabar$.\footnote{Recent experimental 
results~\cite{Abel:2020gbr} indicate that $|\thetabar| \lesssim 10^{-10}$.}
The nucleon EDM can then be determined calculating the 3-point function in 
Eq.~\eqref{eq:corr_edm} in a QCD background with the space-time insertion of the 
topological charge.
The GF allows a definition of the nucleon EDM with no renormalization ambiguities,
and contact terms are avoided due to the finite flow time.
In the right plot of Fig.~\ref{fig:topo_vs_flow}, from Ref.~\cite{Dragos:2019oxn}, 
we show the CP-odd form factor as a function of $Q^2$ 
and we observe that it is independent of the flow time 
for $\sqrt{8t} \simeq 0.6 - 0.7$ fm.
One can then easily perform the continuum limit keeping the flow time 
fixed in physical units.
The resulting neutron EDM~\cite{Dragos:2019oxn} is $|d_n| = 0.00152(71) \thetabar~e$ 
fm, which combined with the 
most recent experimental bounds provides the bound $|\thetabar| < 1.98 \times 10^{-10}$
at the $90\%$ of confidence level. 
In Sec.~\ref{sec:qcedm} we will discuss another application of the GF applied to 
fermion fields that allow the calculation of other CP-violating 
contributions to the nucleon EDM.
\vspace{-0.3cm} 
\subsection{Strong coupling}
\label{ssec:alphas}
A very interesting application of the GF is 
the definition of the strong coupling~\cite{Luscher:2010iy} 
with the expectation value of the density
\be 
\left\langle E(t) \right\rangle = 
\left\langle \frac{1}{4} G_{\mu\nu}^a G_{\mu\nu}^a \right\rangle 
= 
3 \frac{g_0^2}{\left(4 \pi\right)^2 t^2} +O(g_0^4)\,.
\label{eq:energy}
\ee 
From the leading order result we can use 
$E(t)$ to provide a non-perturbative definition of the coupling as 
\be 
\overline{g}^2(\mu) = \left. \frac{16}{3}\pi^2 t^2 \left\langle E(t)\right\rangle\right|_{8t \mu^2=1}\,.
\label{eq:gf_coupling}
\ee
Asymptotic freedom implies that the short flow time behavior of $\left\langle E(t) \right\rangle$
can be described by perturbation theory. 
A perturbative expansion in $D=4-2 \epsilon$ dimensions for a generic SU($N$) gauge theory 
results, for $N_f$ dynamical fermions, in 
\be 
\left\langle E(t) \right\rangle = 
\frac{1}{2}g_0^2\frac{N^2-1}{(8 \pi t)^{D/2}}(D-1)\left[1 + 
c_1 g_0^2 + O(g_0^4)\right]\,,
\ee 
\be 
c_1 = \frac{1}{16\pi^2}(4\pi)^\epsilon(8t)^\epsilon
\left[N\left(\frac{11}{3}\frac{1}{\epsilon} + 
\frac{52}{9} - 3 \ln 3\right) - N_f\left(\frac{2}{3}\frac{1}{\epsilon} + 
\frac{4}{9} - \frac{4}{3} \ln 2 \right) + O(\epsilon)\right]\,.
\label{eq:c1}
\ee 
Renormalizing the bare coupling in the $\MSbar$ removes completely 
the poles in Eq.~\eqref{eq:c1}, leading to the finite expression 
\be 
\left\langle E \right \rangle = 
\frac{3}{4 \pi t^2}\alpha(\mu)\left[1 + k_1 \alpha(\mu) + O(\alpha^2)\right]\,,
\qquad k_1 = 1.0978 + 0.0075 \times N_f\,.
\label{eq:E_ren}
\ee 
The strong coupling $\alpha(\mu)$ is now the renormalized coupling in the 
$\MSbar$ scheme and Eq.~\eqref{eq:E_ren} provides the connection with the 
flow-time dependence of $\left\langle E(t) \right \rangle$, i.e. the GF coupling
defined in Eq.~\eqref{eq:gf_coupling}.
This is the first example of the general result~\cite{Luscher:2011bx}
we have described in Sec.~\ref{sec:gf_gauge}, and it shows 
how to connect correlation functions computed with the GF 
to schemes that are more relevant for phenomenological applications.
We will discuss in Sec.~\ref{sec:gf_ferm} more examples on how to connect
flowed fields with fields at $t=0$ renormalized in the $\MSbar$.
\begin{figure}[h]
     \centering
     \includegraphics[width=6.3cm,clip]{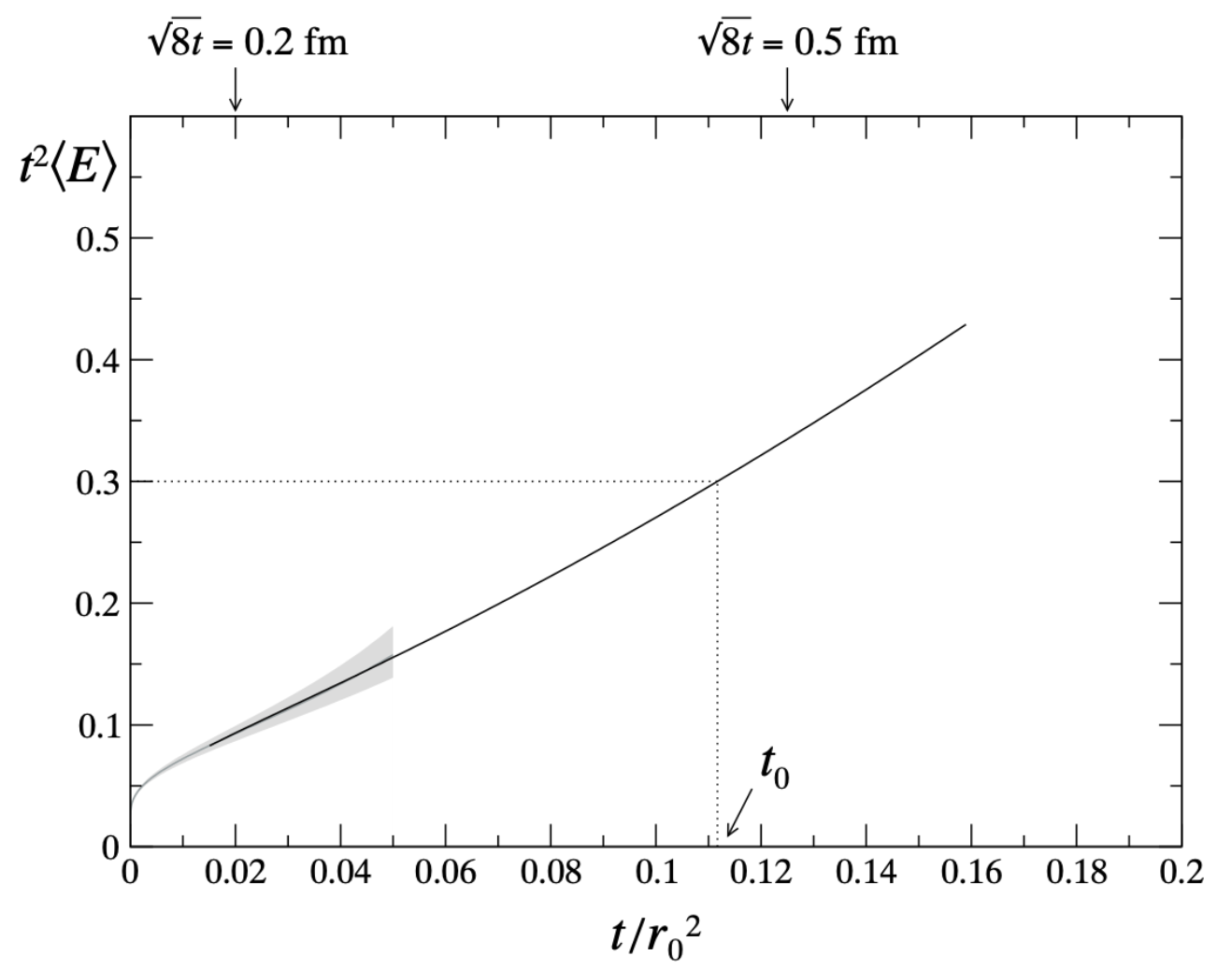}
     \includegraphics[width=6.3cm,clip]{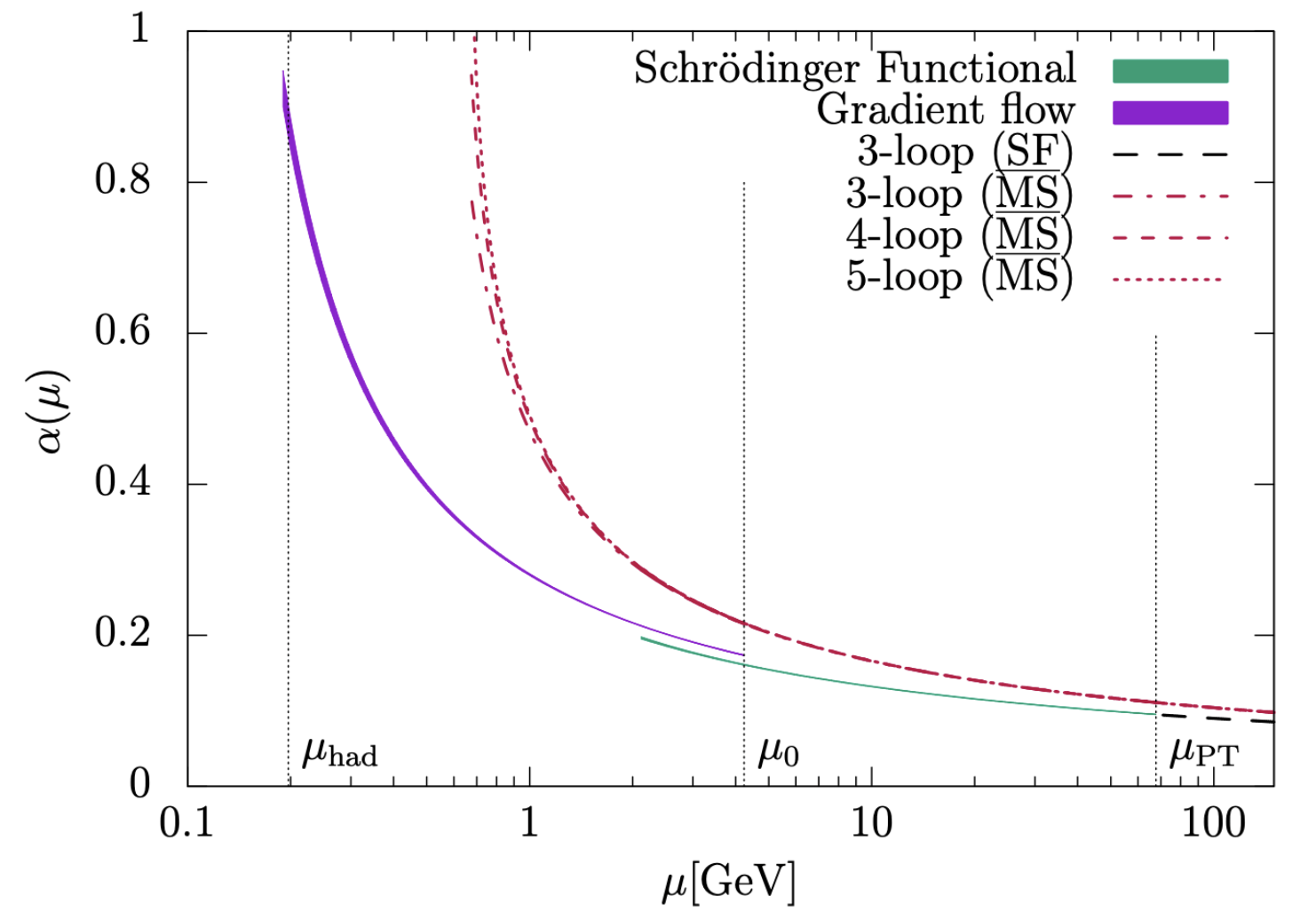}
     \caption{Left plot: 
     Flow-time dependence of the energy $t^2\left \langle E(t) \right\rangle$ 
     in the quenched model. 
     The gray band is the prediction for perturbative QCD (see text) and $t_0$ 
     denotes the flow time value where 
     $t^2 \left. \left\langle E(t) \right\rangle\right|_{t=t_0} = 0.3$.
     Right plot: running coupling as a function of the renormalization scale 
     for $3$ different schemes: the GF scheme described in the main text, 
     the $\MSbar$ scheme,
     and what is denoted as the Schr\"odinger functional scheme.}
     \label{fig:energy_quenched} 
     \vspace{-0.5cm}       
\end{figure}
In the left plot of Fig.~\ref{fig:energy_quenched}, from Ref.~\cite{Luscher:2010iy}, it is shown 
the flow-time dependence of $t^2 \left\langle E \right \rangle$, solid black line,
together with the perturbative estimate\footnote{
     The gray band reflects the uncertainty of 
     the $\Lambda^{\MSbar}$ used for this plot~\cite{Capitani:1998mq}.}
     of Eq.~\eqref{eq:E_ren} 
determined using the $4$-loop
beta function in the $\MSbar$~\cite{vanRitbergen:1997va}. 
We observe a remarkable agreement 
between the perturbative estimate and the non-perturbative lattice data up to 
distances of $\sqrt{8t} \simeq 0.3$ fm corresponding to scales of $\sim 600$ MeV.
In the region $\sqrt{8t} < 0.2$ fm the lattice data are omitted because 
dominated by cutoff effects. 
This is a reflection of a well-known "window" problem encountered
often in lattice QCD calculations. 
In this specific case the problem can be exemplified as follows. 
To properly match the perturbative calculation it is important to be 
at sufficiently short distances, but at the same time we cannot 
go at too small flow times otherwise lattice correlation functions
will be dominated by cutoff effects. 
A well establish technique, 
finite size scaling~\cite{Luscher:1991wu}, provides a robust solution 
for this potential problem.

The calculation of the strong coupling using finite size scaling has 
a long history, and the scheme is usually dubbed as 
Schr\"odinger functional (SF)~\cite{Luscher:1993gh,Jansen:1995ck}, 
because of the choice of boundary conditions in the "temporal" direction. 
For a review see Ref.~\cite{Sommer:2006sj}.
Analogously to the infinite volume definition in Eq.~\eqref{eq:gf_coupling},
the GF coupling with SF boundary conditions~\cite{Fritzsch:2013je} 
can be defined in a small volume calculating the corresponding 
tree-level value of $\left\langle E(t) \right\rangle$ thus leading to 
\be 
\overline{g}_{\mathrm{GF}}^2(L) = \left[\mathcal{N}^{-1}(c,T/L,x_4/T)\cdot t^2 
\left\langle E(t,x_4)\right\rangle\right]_{t=c^2L^2/8}\,,
\ee 
where $\mathcal{N}^{-1}(c,T/L,x_4/T)$ is the tree-level normalization factor 
depending on the temporal $T$ and spatial $L$ size of the box and on the 
time slice where $E(t,x_4)$ is inserted.
In the right plot of Fig.~\ref{fig:energy_quenched},
taken from Ref.~\cite{Bruno:2017gxd},
is shown the running coupling as a function of the renormalization scale 
for $3$ different schemes: the GF scheme we just described, the $\MSbar$ scheme,
and what is denoted the Schr\"odinger functional scheme, where 
the same step-scaling procedure as in the GF scheme is adopted, but the 
coupling is defined from the response of a small variation of a 
specific background field. 
Technical details on the matching between the SF and the GF couplings 
can be found in Refs.~\cite{DallaBrida:2016kgh,Bruno:2017gxd}.
We just note that combining the GF and the SF couplings it is possible 
to cover a wide range of energies and that for the GF coupling 
there are indications~\cite{DallaBrida:2019wur} 
that is more difficult to make contact 
with perturbation theory than with the SF coupling, 
even at $3$-loop.

Other finite volume schemes have been used to define the strong coupling 
using for example periodic boundary conditions for the gauge fields and 
antiperiodic for fermions~\cite{Fodor:2012td}, 
twisted boundary conditions~\cite{Ramos:2014kla} or 
performing an infinite volume limit~\cite{Hasenfratz:2019hpg}.

\vspace{-0.3cm} 
\subsection{Fixing the scale}
\label{ssec:scale}
To convert lattice QCD results obtained in lattice units, 
into dimensionful quantities, one needs to express the 
lattice spacing $a$ in physical units.
This is achieved in hadronic renormalization schemes, 
using dimensionful experimentally measurable quantities.

The GF introduces a new scale, the flow time $t$, thus potentially
providing a new relative way to fix the lattice spacing. 
L\"uscher suggested~\cite{Luscher:2010iy} 
to use the density in Eq.~\ref{eq:energy} at hadronic distances 
to define the scale $t_0$ (see Fig.~\ref{fig:energy_quenched}) by 
$\left. t^2 \left\langle E(t) \right\rangle\right|_{t=t_0} = 0.3
$.
The new scale $t_0$ can be then determined very precisely 
in the continuum limit in terms of the physical quantity of choice,
and it has the advantage that does not need the analysis of correlation
functions at large Euclidean times like for the determination of baryon 
masses.
Even if not measurable 
experimentally $t_0$ allows to compare  
the continuum limit of completely independent calculations expressing
all the lattice results in units of the appropriate power of $t_0$. 
Additionally once $t_0$ is determined 
very precisely in the continuum limit and in physical units, 
it can be used to renormalize other quantities retaining the same level of precision.
A quantity related to $t_0$ has been introduced in~\cite{Borsanyi:2012zs}, 
dubbed $w_0$ which is fixed by the logarithmic derivative with respect to the flow time
$t^2 \left\langle E(t) \right\rangle$.
The quark mass dependence of $t_0$ and $w_0$ is known from chiral 
perturbation theory~\cite{Bar:2013ora}. 
The largest lattice QCD collaborations have nowadays determined very precise values
of $t_0$ and $w_0$ and they are summarized in FLAG 
review~\cite{FlavourLatticeAveragingGroupFLAG:2021npn}.

\vspace{-0.3cm} 
\section{Gradient flow for fermions}
\label{sec:gf_ferm}
Analogously as for gauge fields, it is possible to define a short distance smoothing
operation also for fermion fields. The transformation is not unique,
but we will focus on L\"uscher's proposal~\cite{Luscher:2013cpa}, where 
all the Dirac components are flowed in the same way.
Alternative transformations have been discussed for example in 
Ref.~\cite{Boers:2020lvc}, but they dot seem to provide obvious advantages, neither 
numerically nor theoretically.
The GF for fermions we consider is defined by 
\bea
&&\partial_t \chi(x,t) = \Delta \chi(x,t) \qquad 
\partial_t \bar\chi(x,t) = \bar\chi(x,t) \overleftarrow{\Delta}  \\ 
&&\chi(x,t=0) = \psi(x)\,, \quad \chibar(x,t=0) = \psibar(x) \nonumber
\eea 
where the laplacian $\Delta = D_\mu D_\mu$ is the squared of the 
flowed covariant derivative acting on the fundamental representation
$D_\mu = \partial_\mu + B_\mu$. The gauge field $B_\mu$ has been flowed 
using the GF in Eq.~\eqref{eq:gf_gauge}.

As for the gauge fields a tree-level analysis 
immediately shows the smoothing properties
of the GF over ranges of $\sqrt{8t}$. 
One important difference though 
is that the flowed fermion field requires 
renormalization~\cite{Luscher:2013cpa}, 
$\chi_R = Z_\chi^{-1/2} \chi$.
The GF for fermions is still very powerful because any fermion local field,
such as bilinears, 4-fermion or chromo-electric fields renormalize all multiplicatively
with a renormalization factor depending only on the fermion 
content~\cite{Luscher:2013cpa}. 
If a local field $\mcO$ contains $n$ fermion and anti-fermion fields,
the field will renormalize as $\mcO_R = Z_\chi^{-n/2} \mcO$. This is a consequence
of the absence of short-distance singularities once we define the local field with 
an operator product expansion. 
A prime example of this phenomenon is the renormalization
of the scalar density. In lattice QCD the scalar density $\psibar \psi$
mixes with lower dimensional operators generating $1/a^3$ power divergences, 
if the lattice action breaks chiral symmetry, or $1/a^2$ when using Ginsparg-Wilson 
fermions. The flowed scalar density 
$\Sigma(t) = \left\langle \chibar(t) \chi(t) \right\rangle$ 
instead renormalizes multiplicatively, i.e. $\Sigma_R(t) = Z_\chi^{-1} \Sigma(t)$.
Considering how much the lattice community has struggled over power divergences,
this result is quite remarkable and can be applied to many phenomenologically relevant
flavor observables. 
Once the renormalization of the local fields is resolved, it remains to connect the 
renormalized matrix elements of flowed fields to the physical ones at $t=0$. 
The connection can be done in $2$ ways: 
using Ward identities~\cite{Luscher:2013cpa,DelDebbio:2013zaa,Shindler:2013bia} 
or relying on an operator product expansions at short 
flow time~\cite{Luscher:2013vga}, also called short flow-time expansion (SFtX).

Ward identities (WI) involving flowed fields are different from the standard WI.
Using the modified modified chiral WI~\cite{Luscher:2013cpa,Shindler:2013bia}
and the spectral decomposition of pseudoscalar
$P(x)$ 2-point functions it is possible to derive the following non-perturbative 
identity for the renormalized chiral condensate $\Sigma_R$
\be 
\Sigma_R = Z_P \lim_{m_R \rightarrow 0} \frac{G_\pi}{G_\pi(t)} \Sigma(t)\,,
\ee
where $G_\pi=\left\langle 0| P | \pi \right\rangle$ and 
$G_\pi(t) = \left\langle 0| P(t) | \pi \right\rangle$, while 
$\Sigma(t)=\left\langle \chibar(x,t) \chi(x,t)\right\rangle$.
The problem of the power divergence is resolved and the condensate is calculable
only knowing $Z_P$, $2$-point pseudoscalar densities 
and the flowed scalar expectation value.
All these results are formally obtained in the continuum. 
At finite lattice spacing one should still make sure 
to avoid to use flow times too close to the values 
of the lattice spacing in use. After performing the continuum limit 
at fixed value of the flow time in physical units,
the final result for $\Sigma_R$ should be flow-time independent.
\vspace{-0.3cm} 
\subsection{Short flow-time expansion}
\label{ssec:sftx}
It is not always possible to use WI to connect flowed renormalized 
fields, and the corresponding matrix elements, with the physical ones at $t=0$.
It is possible though to use an OPE at small flow time, also known 
as short flow-time expansion (SFtX).
The SFtX can be exemplified as follows. 
On the lattice we compute the correlation function of interest,
where the local field $\mcO_{i,R}(t)$ 
is flowed and renormalized. The only renormalization needed is for the 
flowed fermion fields, beside the coupling and the quark mass.
There are several ways to achieve this. One way is to use the following 
regularization independent and gauge invariant condition
\be 
\left\langle \mathring{\bar\chi}(x;t) \overleftrightarrow {\slashed D} 
\mathring\chi(x;t)\right\rangle = - \frac{2 N N_f}{(4\pi)^2 t^2}\,.
\label{eq:ringed_np}
\ee 
The $\mathring\chi$ and $\mathring{\bar\chi}$ 
are the so-called "ringed" fields~\cite{Makino:2014wca,Makino:2014taa} 
which are now free from UV divergences.
Another possibility would be to normalize the correlation functions 
with vector $2$-point functions~\cite{Hasenfratz:2022wll}.
Independently of the method used the renormalization is greatly simplified,
as there are no power divergences and no mixing with other fields, i.e.
the renormalization is multiplicative.

Having renormalized the flowed fields one can use the SFtX 
\be
\mathcal{O}_{i,R}(t) \underset{t \rightarrow 0}{\sim}\sum_i 
c_{ij}(t,\mu) \mathcal{O}_{j,R}(t=0,\mu)\,,
\ee
to determine the target renormalized matrix element at $t=0$.
The renormalization problem is now shifted to the determination of the 
matching coefficients $c_{ij}(t,\mu)$. The renormalized matrix elements 
$\mathcal{O}_{j,R}(t=0,\mu)$ are then determined in the same renormalization 
scheme used to determine the matching coefficients. 
When the SFtX receives contributions from lower dimensional operators 
the matching coefficients need to be determined non-perturbatively 
on the lattice~\cite{Maiani:1991az,Kim:2021qae,Mereghetti:2021nkt}. 
In Sec.~\ref{sec:qcedm} we discuss a method we have 
devised to determine non-perturbatively 
the matching coefficients of the power divergences.  
For contributions from fields of the same dimensions one can determine 
the matching coefficients directly in the continuum in perturbative QCD.
This is a great simplification, compared with the 
original renormalization problem. 
Moreover the matching coefficients can be determined using standard techniques,
where the SFtX is inserted in off-shell amputated and gauge-fixed 
1PI correlation functions. The matching coefficients, after the renormalization procedure,
will be independent on the gauge-fixing procedure and on the specific probes 
used to determine them.
\vspace{-0.3cm} 
\section{Quark-chromo electric dipole moment}
\label{sec:qcedm}
At hadronic scales BSM CP-violating sources for a non-vanishing EDM 
are encoded in higher-dimensional
effective operators, containing the degrees of freedom of QCD and QED.
One of these operators is the quark-chromo EDM (qCEDM) 
\be 
\mathcal{O}_{\textrm{CE}} (x)= 
\sum_{f=u,d,s,\ldots} \overline{\psi}_f(x) \gamma_5 \sigma_{\mu\nu}
G^a_{\mu\nu} T^a \psi_f(x)\,.
\ee
In order to interpret future positive or null EDM 
experiments, and thus disentangle all possible CP-violating sources, 
it is important to determine renormalized qCEDM hadronic matrix elements
using lattice QCD.
The challenge of these type of calculations 
has been the very complicated renormalization
pattern when using momentum-subtraction 
schemes~\cite{Bhattacharya:2015rsa,Cirigliano:2020msr}.
We have proposed~\cite{Kim:2018rce,Rizik:2020naq,Kim:2021qae,Mereghetti:2021nkt} 
to use the GF to resolve the renormalization of the qCEDM and other CP-violating fields.
We use the qCEDM as an example of the renormalization procedure using the SFtX 
we presented in the previous 
section\footnote{An alternative method for the renormalization of the 
qCEDM has been proposed~\cite{Izubuchi:2020ngl}
based on coordinate-space renormalization.}.

\begin{figure}[t]
     \centering
     \includegraphics[width=6.3cm,clip]{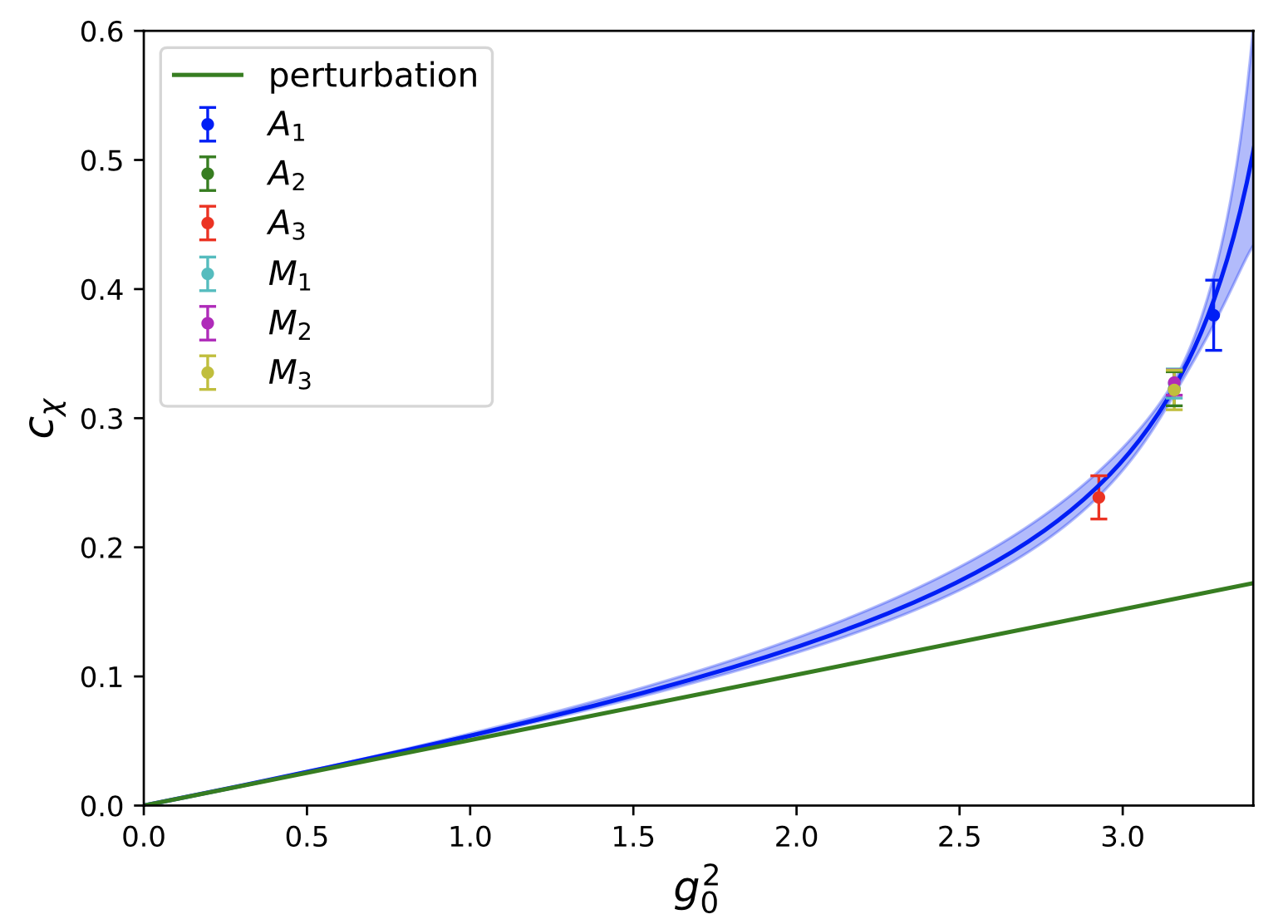}
     \includegraphics[width=6.3cm,clip]{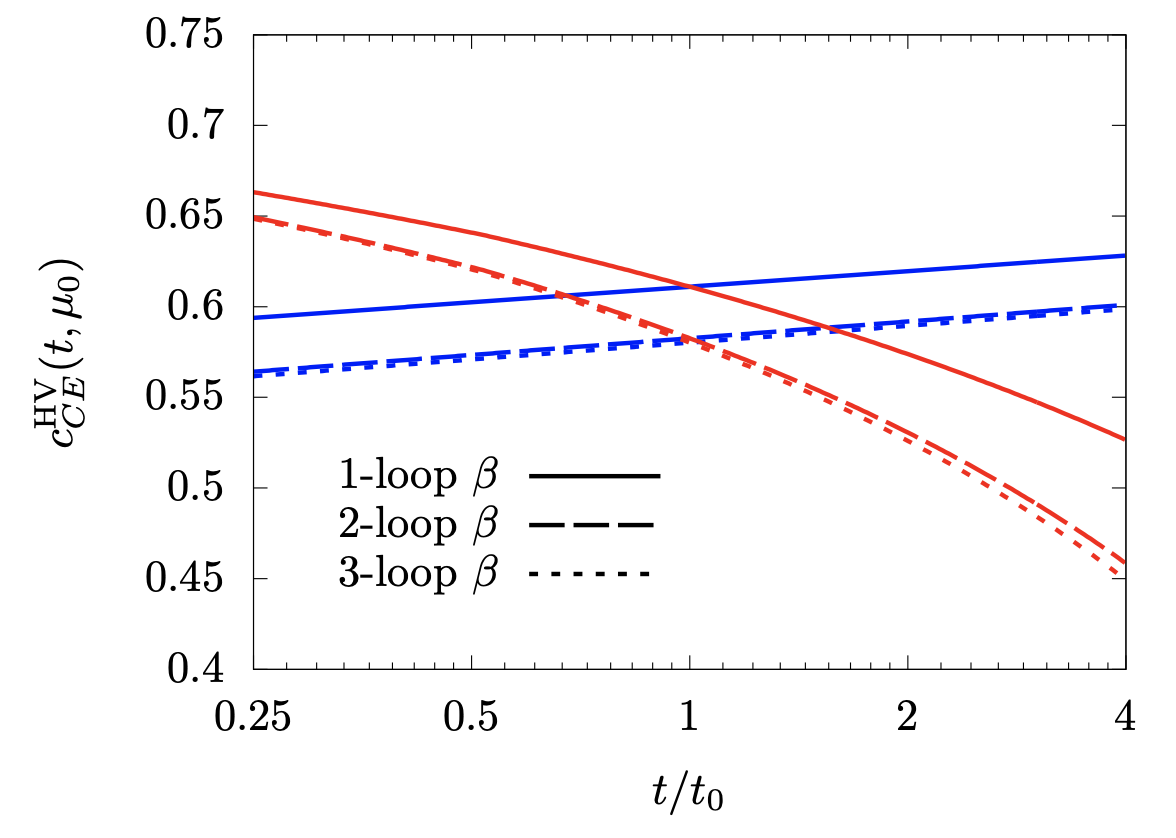}
     \caption{Left plot: bare coupling dependence of $c_\chi = t Z_\chi c_P(t,\mu)$. 
     Lattice data with different colors (see Ref.~\cite{Kim:2021qae}) are described 
     with a Pade' approximant constrained at small coupling by 
     perturbation theory~\cite{Rizik:2020naq}. 
     Right plot: scale dependence of the qCEDM matching 
     coefficient. See main text and Ref.~\cite{Mereghetti:2021nkt} for details 
     on the plot.}
     \label{fig:power_symlat}
     \vspace{-0.7cm}        
\end{figure}

For simplicity for now we consider the single flavor qCEDM.
We now imagine that the flowed qCEDM has been defined with "ringed" fields,
i.e. it has been renormalized, 
\be 
{\mathcal{O}}_{CE}^R(x;t) = 
\mathring{\bar\chi}(x;t) \sigma_{\mu\nu} G_{\mu\nu}(x;t)\mathring\chi(x;t)\,.
\label{eq:qcedm_ring}
\ee
The leading contribution in the SFtX comes from the pseudoscalar density, 
that generates a power divergence. While in a typical lattice scheme this would 
be a power divergence in $1/a^2$ in the SFtX is a power divergence in $1/t$ 
and the coefficient, $c_P(t,\mu)$, should and can be determined non-perturbatively.
In Ref.~\cite{Kim:2021qae} we have devised a gauge-invariant non-perturbative 
method to determine the matching coefficient that relies on the calculation 
of the ratio
\be 
\left[R_P(x_4;t)\right]_R = 
t \frac{\left[\Gamma_{CP}(x_4;t)\right]_R}{\left[\Gamma_{PP}(x_4)\right]_R}\,,
\label{eq:R}
\ee 
\be 
\Gamma_{CP}(x_4;t) = 
a^3\sum_{\bf{x}}\left\langle{\mathcal{O}}_{CE}^{ij}(x_4,\bx;t)
P^{ji}(0,\bzero;0)\right\rangle\,, \quad
\Gamma_{PP}(x_4) = 
a^3\sum_{\bx}\left\langle P^{ij}(x_4,\bx) P^{ji}(0,\bzero)\right\rangle\,.
\label{eq:gamma_CP}
\ee 
Once we define the qCEDM in terms of the "ringed" fields and we renormalize 
the pseudoscalar density, the ratio~\eqref{eq:R} has a well defined continuum 
limit at fixed renormalized coupling and it can be used to determine the 
matching coefficient of the qCEDM into the pseudoscalar density, $c_P(t,\mu)$. 
In Ref.~\cite{Kim:2021qae}, we did not have at our disposal
a complete determination of the "ringed" field, so we studied the 
dependence on the bare coupling of $ c_\chi = t Z_\chi c_P(t,\mu)$,
shown in the left plot of Fig.~\ref{fig:power_symlat}.
The colored data points represent lattice data at $3$ different lattice 
spacings and $3$ different pion masses (see Ref.~\cite{Kim:2021qae} for details)
and the green line represents the first order perturbative result for 
$c_P(t,\mu)$~\cite{Rizik:2020naq}. A Pade' approximant, constraint 
with our perturbative result, describes well the 
data showing the relevance of a perturbative 
calculation of the matching coefficients, even for power divergences.
In general the matching coefficient will depend on the scheme $S$ 
used to determine $Z_P^{(S)}$ 
and on the definition of "ringed" fields. When multiplied with the renormalized 
matrix element in the SFtX the dependence on the scheme $S$ will cancel out 
and the subtracted operator $\mcO_{\mathrm{sub}}(x;t) = \mcO^R_{CE}(x;t) - 
c_P^S(t,\mu) P_R^S(x;\mu)$ will have a SFtX with contributions only from 
fields of the same or higher dimensions.
To calculate the missing matching coefficients one can rely on perturbative 
QCD. We setup the theory in $D=4-2\epsilon$ dimensions and impose the 
following matching conditions on amputated 1PI correlation functions
\be 
Z_\chi^{-n/2} \left\langle\left(\psi\right)^{n_\psi} \left(\bar\psi\right)^{n_{\bar\psi}} 
\left(A_\mu\right)^{n_A}\mathcal{O}_i(t)\right\rangle^\text{amp} = 
c_{ij}(t,\mu) \, \left(Z_{jk}^\text{MS}\right)^{-1}\left\langle\left(\psi\right)^{n_\psi} 
\left(\bar\psi\right)^{n_{\bar\psi}} \left(A_\mu\right)^{n_A} 
{\mathcal{O}}_k \right\rangle^\text{amp}\,, 
\label{eq:matching_eq}
\ee 
where $n$ is the total number of flowed fermion fields and the number of 
external fermion or gluon fields is chosen depending on what matching coefficient
we want to isolate. 
The renormalized flowed fermion fields are given by 
\be 
\chi_R(x;t) = (8\pi t)^{\varepsilon/2} \zeta_\chi^{1/2} \mathring\chi(x;t)\,, \quad 
\bar\chi_R(x;t) = (8\pi t)^{\varepsilon/2} \zeta_\chi^{1/2} \mathring{\bar\chi}(x;t)\,,
\label{eq:ringed_norm}
\ee 
where 
$\zeta_\chi$ has been determined at 1-loop in~\cite{Makino:2014taa,Makino:2014wca}
and up to 2-loops in Refs.~\cite{Harlander:2018zpi,Artz:2019bpr}.

The calculation of the matching coefficients can be summarized with the 
following steps. First renormalize the flowed fermion fields, e.g. using 
"ringed" fields, then expand in powers of external scales. The $t=0$ Feynman loop 
integrals vanish, being scaleless or, in other words, reflecting the cancellation 
between UV and IR poles. The $t >0$ Feynman loop integrals 
are only IR divergent and we regulate them in dimensional regularization. 
The resulting poles should match the UV poles in the $t=0$ side, thus can be removed 
with the renormalization needed at $t=0$. One is then left with a finite 
matching coefficient obtained in the same scheme used for the renormalization.
In this way it is sufficient to just calculate the correlation 
functions containing the flow fields,
expand in the external scale and then remove the IR divergences with the 
renormalization factors in the scheme we want to evaluate the matching coefficients.
For full details of the calculation I refer to Ref.~\cite{Mereghetti:2021nkt}.
As an example the matching coefficient of the qCEDM into itself is given by 
\be
c_{CE}(t,\mu) = \zeta_\chi^{-1} + 
\frac{\alpha_s}{4\pi} \left[ 2(C_F - C_A) \log(8\pi\mu^2 t) - 
\frac{1}{2} \Big( (4+5 \delta_\mathrm{HV}) C_A + 
(3- 4\delta_\mathrm{HV} ) C_F \Big) \right]
\ee
where it is explicit the renormalization of the flowed fermion field with $\zeta_\chi$ and 
$\delta_\mathrm{HV}=0$ or $1$, whether we define $\gamma_5 \sigma_{\mu\nu}$ in naive dimensional
regularization or in the 't Hooft-Veltman~\cite{tHooft:1972tcz,Breitenlohner:1977hr} 
scheme.
To estimate the uncertainty of the matching coefficient stemming from the truncation 
of the perturbative expansion we vary $c_{CE}(t,\mu)$ around the 
scale $t_0 = \frac{1}{8 \pi \mu_0^2}$ (not to be confused with $t_0$ discussed in 
Sec.~\ref{ssec:scale}) with $\mu_0=1.13$ GeV the scale in the MS scheme, corresponding 
to $\overline{\mu}_0=3$ GeV in the $\MSbar$ scheme. 
In the right plot of Fig.~\ref{fig:power_symlat} we show the scale dependence of 
$c_{CE}(t,\mu)$ in the t'Hooft-Veltman scheme where we keep the 
running coupling fixed at the scale $\mu_0$ (blue curves) or we vary it 
together with the matching coefficient (red curves). The spread of the curves,
around $10-20 \%$, 
gives us an indication of the uncertainty due to the truncation of the expansion.
It is a rather strong indication that it would be beneficial to extend the 
calculation to $2-$loops and work in this direction is in progress.
\vspace{-0.4cm} 
\section{Other applications}
\vspace{-0.1cm}
\subsection{Finite temperature}
In a hot QCD medium, transport coefficients characterize the motion 
of heavy quarks. They can be determined performing an heavy quark mass, $M$, 
expansion in $T/M$, where $T$ is the temperature. The first $2$ leading terms
of the expansions can be estimated calculating 2-point static quarks 
correlation functions of chromo-electric and chromo-magnetic fields.
Results for the momentum diffusion coefficient, $\kappa$, have been 
obtained using the GF~\cite{Altenkort:2020fgs,Brambilla:2022xbd}.
The main reason for the use of the GF for these type of calculations 
is to improve the signal-to-noise ratio especially at large Euclidean times.
While these type of studies are still in their infancy, they can pave the 
way for a new application of the GF, especially because 
fields contributing to the heavy quark expansion need to be 
renormalized non-perturbatively.
\vspace{-0.3cm}
\subsection{Other results for fermion local fields}
One of the important application of the GF is the 
renormalization of local fermion fields, like the qCEDM, as discussed in 
Secs.~\ref{sec:gf_ferm} and \ref{sec:qcedm}. 
The matching coefficients of fermion bilinears have been 
calculated for example in Refs.~\cite{Hieda:2016lly}.
Other local fermion fields that 
can benefit from the GF are $4$-fermion operators. 
Perturbative calculations of the matching 
coefficients for the $4$-quark operators, relevant for $B_K$
and other contributions to the effective electroweak Hamiltonians, have been 
performed up to $2$-loops~\cite{Suzuki:2020zue,Harlander:2022tgk}.
The short-distance behavior of the product of $2$ fermion 
bilinears can be described by an OPE. The renormalization of the 
operators contributing to the OPE can be performed with the GF and the 
corresponding matching coefficients, up to and including 
$D=4$ fields have been calculated at $2$-loops in Ref.~\cite{Harlander:2020duo}.
\vspace{-0.3cm}
\subsection{Energy-momentum tensor}
The renormalization of the energy-momentum tensor on the lattice is 
a rather non-trivial task. The trace of the tensor mixes with the identity
and if we include fermions there is a mixing under renormalization 
between $5$ different dimension 4 fields. 
Using a SFtX, in refs.~\cite{Suzuki:2013gza,Makino:2014taa} the matching coefficients 
in the pure gauge theory and with the inclusion of fermions have been 
determined. This method has been successfully tested numerically at finite 
temperature in the SU($3$) gauge theory in Ref.~\cite{Kitazawa:2016dsl} 
and with $N_f=2+1$ dynamical fermions in~\cite{Taniguchi:2016ofw} 
with $1-$loop matching\footnote{In Ref.~\cite{Taniguchi:2016ofw} 
results for the chiral condensate at finite temperature have been 
presented following the methods described in Sec.~\ref{sec:gf_ferm}.},
and in Ref.~\cite{Taniguchi:2020mgg} with a $2$-loop matching,
using the perturbative calculation of Ref.~\cite{Harlander:2018zpi}.
A different approach has been proposed in 
Ref.~\cite{DelDebbio:2013zaa} based on WI imposed to recover 
space-time symmetries at finite lattice spacing. It would be 
important to test numerically this approach.
\vspace{-0.3cm}
\subsection{Non-perturbative renormalization scheme}
Even though the GF is not a renormalization group transformation, 
it can be used to define non-perturbative renormalization schemes.
Attempts in this direction have started following different strategies
in Refs.~\cite{Monahan:2013lwa,Hasenfratz:2022wll,Battelli:2022kbe}.

\vspace{-0.3cm}
\section{Conclusive remarks and acknowledgements}
The gradient flow in recent years has played a prominent role in improving
several aspects of lattice QCD calculations. It has provided a
renormalizable definition of the topological charge, a new dimensionful scale
to set the lattice spacing in physical units and an alternative definition 
of the strong coupling. It has also provided a new method to resolve 
challenges related to the renormalization of local fields, especially 
in the presence of power divergences and complicated mixing.
Continuing to study all the nuances of the gradient flow, and extending 
the perturbative calculations of matching coefficients is 
a necessary step to address old and new challenges: e.g. in the 
non-perturbative studies of the CP-violating sources 
of the electric dipole moment, the effective electroweak Hamilonian 
and parton distribution functions.

I want to thank my collaborators N. Brambilla, A. Hasenfratz, R. Harlander, 
J. Kim, Z. Kordov, V. Leino, T. Luu, E. Mereghetti, C. Monahan, G. Pederiva, 
M. Rizik, P. Stoffer, A. Vairo, X. Wang, O. Witzel for many interesting 
discussions about the gradient flow and most enjoyable collaborations.

\bibliography{ref}
\vspace{-0.3cm}

\end{document}